\documentclass[twocolumn,showpacs,preprintnumbers,amsmath,amssymb,aps,pra]{revtex4}
\usepackage{graphicx}
\usepackage{dcolumn}
\usepackage{bm}
\def\bfr{{\bf r}}
\def\nab{{\bf \nabla}}
\begin{document}
\title{Condensate wave function of neutral Bose atoms in power-law traps}
\author{Sang-Hoon Kim}
\email{shkim@mail.mmu.ac.kr}
\affiliation{Division of Liberal Arts, 
Mokpo National Maritime University, Mokpo 530-729, {\rm and}
 Institute for Condensed Matter Theory, Chonnam National University,
Gwangju 500-757, Korea}
\author{Chang Sub Kim}
\affiliation{Department of Physics {\rm and}
Institute for Condensed Matter Theory, Chonnam National University,
Gwangju 500-757, Korea}
\date{\today}
\begin{abstract}
The macroscopic quantum states of a condensed neutral Bose gas
in one-dimensional power-law traps are obtained by solving
the Gross-Pitaevskii equation numerically.
A suitable candidate for a trial wave function for 
the variational calculation of ground state energy
is  suggested, and shown that the suggested function
produces a lower ground state energy 
than the conventional Gaussian trial wave function.\\
\end{abstract}  
\maketitle

Bose-Einstein condensation (BEC) is a first order 
phase transition of bosonic particles  
when the thermal de Broglie wavelength exceeds the
mean spacing between particles.
The first BEC was observed
 in a series of the landmark experiments
 in 1995  on alkali vapors  in which the atoms 
were confined in magnetic traps and cooled down to 
sub $\mu$-Kelvin temperatures.
Some of the relevant features of these trapped Bose gases
are that they are inhomogeneous, and
 a harmonic external potential 
 is typically used  to describe such a trapped system.
However, due to the development of the experimental 
techniques of magnetic traps,  
the shape of the external trap can be managed intentionally,
and power-law traps in the form of $V_{trap} \sim r^k  (k>0)$
are accessible now.

Bosonic atoms in optical lattices and periodic potentials
often require non-harmonic traps 
such as elliptic or a trough type traps \cite{bron,gonz}.
Therefore, it is necessary to predict the condensate wave functions
in arbitray power traps since it has related with theoretical study
of BEC such as the variational calculation of the ground state energy
of the condensed atoms.
In harmonic traps a Gaussian trial function of
$exp(-r^2)$ is generally applied to the variational calculation, 
but the form is not guaranteed the adequate low energy
 in the power-law traps.
Surprisingly, the shape of the condensate function
 at power-law traps has not been investigated in spite of
many studies of power-law traps\cite{bagn,bayi}.
Most of the previous works reported mainly 
transition temperatures and condensate fractions.

At zero temperature, the condensate wave function of the 
typical model can be described by a non-linear Schr\"{o}dinger 
equation, named as the Gross-Pitaevskii equation (GPE).
 It is a mean-field approximation for the macroscopic 
wave function of weakly interacting bosons.
In this short note we'll solve the one-dimensional
GPE at a few power-law traps numerically, 
and suggest the shapes of the new 
condensate wave functions that lowers 
the ground state energy than
the Gaussian form.
Then, we will follow the three steps to compare 
the three energies:
1) Apply the above numerical wave functions 
 to the Hamiltonian directly and obtain the ground state
energy numerically.
2) Apply the new trial function to the Hamiltonian,
and obtain the lowest state energy analytically.
3) Apply the conventional Gaussian trial function
to the hamiltonian and obtain the lowest state energy analytically.

The GPE for $N$ interacting bosons 
confined by an external trap potential $V_{trap}$ 
and under delta-type interaction is written as
\begin{equation}
i \hbar \frac{\partial }{\partial t}\Psi(\bfr,t) = 
\left\{-\frac{\hbar^2}{2m} \nab^2 + V_{trap}(\bfr)
+ g |\Psi(\bfr,t)|^2  \right\}\Psi(\bfr,t),
\label{8}
\end{equation}
where $g$ is the interaction strength given by
$g=4\pi\hbar^2 a/m$ \cite{park,dalf1}.  
The $a$ is the s-wave scattering length and $m$ is the mass.
The $\Psi(\bfr,t)$ is real and normalized to 
the total number of particles, $N$. 
We choose the power-law potential for the external trap
 of the form $ V_{trap}(\bfr) = c r^k $
where $c$ is a positive constant that satisfies the dimensionality
condition, and $k= 1, 2, 3, ...$. 
Note that $k=1$ corresponds to a linear trap,
and $k=4$ to a quadratic trap, and so on.

A scaling of variables such as  
$ a_{ho}= \sqrt{\hbar/m\omega} $ for the length,
and $\hbar \omega$ for the energy
 reduces Eq. (\ref{8}) to a dimensionless form \cite{park,dalf1}.
It corresponds to $\hbar = m = \omega=1$ in Eq. (\ref{8}).
We consider the one dimensional case
because it is good enough for the comparison.
\begin{equation}
i \frac{\partial \psi}{\partial t} =
\left[ -\frac{1}{2 }\frac{\partial^2}{\partial x^2} 
+ \frac{1}{2} {x}^k + P |\psi|^2
\right] \psi (x,t), 
\label{12}
\end{equation}
 For convenience we chose  $c=1/2$
with an adjusted dimension of energy.
The  $P$ is the dimensionless number
parameter defined by 
$ P = 4\pi N |a| /a_{ho}, $
which is  the only variable that differentiate each atoms.
Note that  $|a| /a_{ho}$ is order of $10^{-3}$
for alkali atoms, and 
then, P=1 corresponds approximately to $N=10^2$.
 $\psi (x,t)=\psi(x) exp(-i\mu t/\hbar)$  
is normalized to unity,
and $\mu$ is the chemical potential.

\begin{figure}
\includegraphics{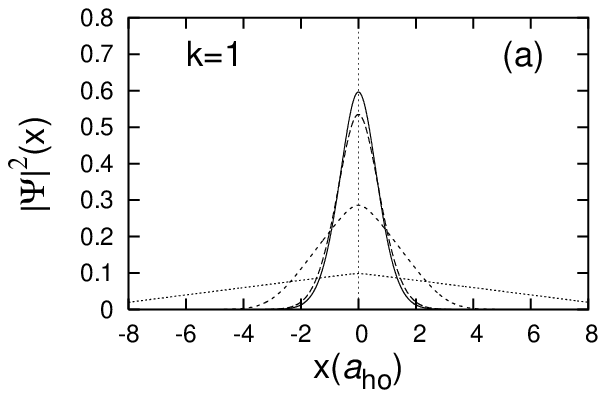}\\
\includegraphics{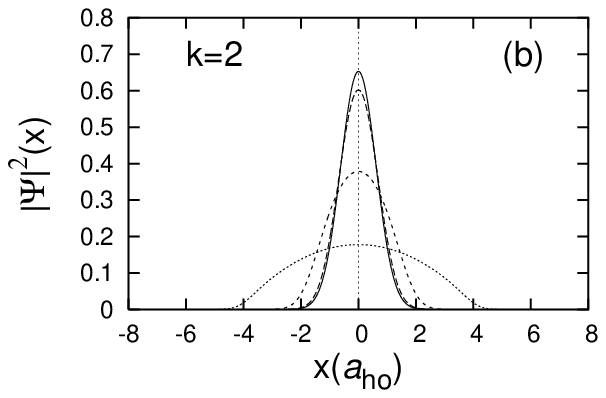}\\
\includegraphics{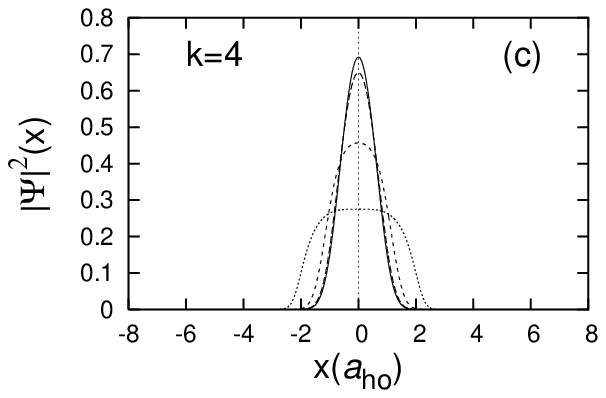}
\caption{
Density profile of condensed bosons 
 at one-dimensional power-law traps.
(a) for linear trap, $k=1$,  (b) for harmonic trap,  $k=2$,
 and (c) for quadratic trap,  $k=4$.
Solid line for $P=0.1$, long dashed lines for $P=1$,
short dashed line for $p=10$, and dotted lines for $P=100$.}
\end{figure}

We have carried out a numerical procedure 
for Eq. (\ref{12}),
and have obtained the static solution $\psi(x)$
 at various powers.
Fig. 1(a) is for the linear trap ($k=1$),  
and Fig. 1(b) is
  for the well-known harmonic trap ($k=2$),
 and Fig. 1(c) is for the quadratic trap ($k=4$).
For small number of particles there 
is little dependence of the traps, 
and the wave function is close to the Gaussian.
However, for large number of particles
 the wave function does not resemble
the Gaussian any more,
 and it depends  seriously on the power of the traps.

In the strongly repulsive limit of $P \gg 1$, 
the kinetic energy term in Eq. (\ref{12}) is neglected 
and then the solution becomes 
$\psi^2 (x) = (\mu - \frac{1}{2}x^k)/P.$
However, it is 
not applicable for most of large realistic $P$'s.
Therefore, we need to propose an alternative trial function
 for a large number of particles.
We  suggest  it as $\psi(x) \sim  exp(- x^k)$ 
 at the $k$-th power-law traps
where $k= 1, 2, 3, ...$.
Then, the Gaussian trial function
 is a special case for the harmonic traps.

\begin{figure}
\includegraphics{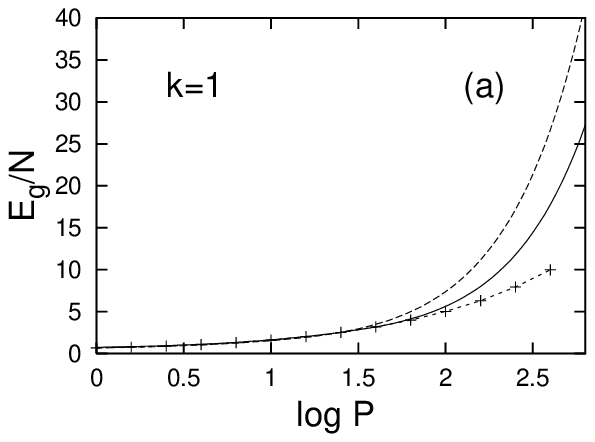}\\
\includegraphics{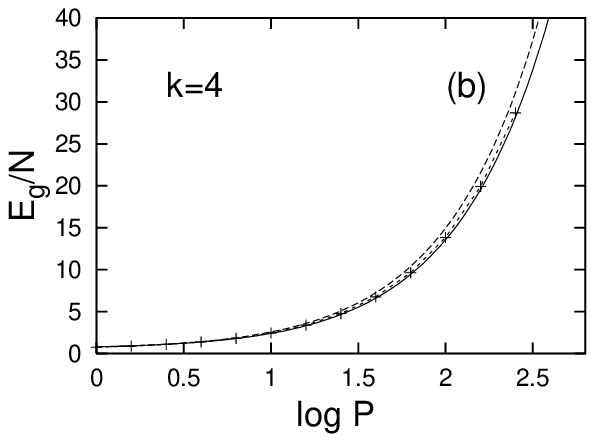}\\
\caption{
The lowest energy per particles.
(a) $k=1$, (b) $k=4$.
Solid line from the new trial function, 
dashed line from the Gaussian trial function,
and cross with dotted line from the numerical calculation.}
\end{figure}

To check the effectiveness of the new trial function,
we need to compare the ground state energy densities
produced by the new ones with two others:
One from the direct numerical calculation
by the Fig. 1, and the other from the analytic
calculation by the conventional Gaussian
trial functions.
Similar to Eq. (\ref{12}), the ground state energy per particle
at one-dimension
can be written in the dimensionless form \cite{dalf,dodd}
\begin{equation}
\frac{E}{N} = \frac{1}{2}\int_{-\infty}^{\infty} dx
 \left[  \left( \frac{\partial \psi}{\partial x} \right)^2 
+ x^k \psi^2(x) + P\psi^4(x) \right].
\label{36}
\end{equation}

One can minimize the energy densities in Eq. (\ref{36}) 
using the variational {\it ansatz} with the new trial function
\begin{equation}
\psi_K(x) = A_k e^{-\frac{1}{2}\left(\frac{x}{d}\right)^k},
\label{38}
\end{equation}
and with the Gaussian trial function
\begin{equation}
\psi_G(x) = A_2 e^{-\frac{1}{2}\left(\frac{x}{d}\right)^2}.
\label{39}
\end{equation}
The $d$ is the variational parameter which fixes 
the width of the condensate.
$A_k$ is determined from the constraint 
$\int_{-\infty}^{\infty} \psi^2(x) dx = 1$. Then,
$A_k = \sqrt{ k/\{4\pi \Gamma(3/k) d^3\}}$.

Substituting $\psi_{K,G}$ in Eqs. (\ref{38}) and (\ref{39})
 into Eq. (\ref{36}), we obtain the energies per particle
$E_{K,G}/N=f_{K,G}$ respectively as
\begin{equation}
f_K(d) = \frac{(k+1)\Gamma\left(\frac{1}{k}\right)}
{8\Gamma\left(\frac{3}{k}\right) d^2}
+ \frac{3}{2 k} d^k
+\frac{kP}{ 2^{3+\frac{3}{k}}\pi \Gamma \left(\frac{3}{k}\right) d^3},
\label{41}
\end{equation}
and
\begin{equation}
f_G(d) = \frac{3}{4 d^2} +\frac{\Gamma\left( \frac{k+3}{2} \right)}
{\sqrt{\pi}} d^k +\frac{P}{4\pi\sqrt{2\pi}d^3 },
\label{42}
\end{equation}
where $\Gamma$ is the Gamma function. 
It is clear that $f_K = f_G$ when  $k=2$.

We have determined the minimum of the energy densities
 by solving $df_{K,G}/dd |_{d_c} = 0$ numerically, and
 plotted $f_{K,G}(d_c)$   as a function of $\log P$.
Fig. 2(a) is for the linear trap, and Fig. 2(b) is
 for the quadratic trap.
At small number of atoms such as $\log P < 1$,
the lowest energies by the two trial functions
show similar values
with the numerical calculation.
On the other hand, for large number of atoms
the trial function does not always 
guarantee an effective approximation
for the ground state.
It depends strongly on the power of the trap.
At the quadratic trap
 the two trial functions are still good enough
for the ground state even for large number of atoms,
 but at  the linear trap it is not effective anymore.
However, as we see, the new trial functions 
always produce the lower energies
 than the Gaussian. 
 
We studied the solution of the one-dimensional GPE
of neutral Bose atoms under power-law traps
numerically, and suggested 
 a better trial wave function,  $exp(-r^k)$, 
 beyond Gaussian.
We showed that the new trial function 
produce lower energy densities 
than the Gaussian.
It can be applicable effectively  
 for condensed wave fucntions at any non-harmonic traps.

We send our special thanks to professors Q-Han park for this research.
This work was supported by the special research fund of 
Chonnam National University in 2004.

\end{document}